\documentclass[aps,twocolumn,amsmath,prl]{revtex4-1}
\usepackage{graphicx}
\usepackage{subfigure}
\usepackage{amsmath}
\usepackage{bibentry}

\begin{document}

\title{The topological vacuum in exceptional Brillouin zone}

\author{Ye Xiong} 
\email{xiongye@njnu.edu.cn}
\affiliation{Department of Physics and Institute of Theoretical Physics
  , Nanjing Normal University, Nanjing 210023,
P. R. China}

\begin{abstract}
	The vacuum of a band system, with respect to particles or holes, can become
	topologically nontrivial when the exceptional points of the non-hermitian
	Hamiltonian spread over the whole Brillouin zone. The coalescence of the
	eigenstates eliminates the geometric multiplicity of the
	Hamiltonian and puts the remaining bands on a topological background. We
	find several distinct phenomena in these systems: a single band
	topological model, a two-band Chern insulator in which the topological number of
	the low band does not compensate to that of the up band, a topologically protected
	boundary state that is less damping when stronger dissipative forces are subjected and
	topologically protected in-gap extended states. Our results are presented by the
	vibrational spectrum on a classical lattice. The active constrains are the key
	ingredient to collapse the demanding Hilbert subspace.   
\end{abstract}

\maketitle

{\it Introduction.---} In condensed matter physics, a band must be topologically trivial
when its project operator, $\hat P_n = \sum_k |n(k)\rangle \langle n(k)|$, commutes with
the translation operator in the reciprocal space, $e^{I\hat r \cdot \delta k}$, where
$|n(k)\rangle$ is the eigenstate of the $n$th band, $\hat r$ is the position operator and
$\delta k$ is an infinitesimal displacement along the Wilson loops in the Brillouin zone
(BZ)\cite{Yu2011a, Soluyanov2011, Kivelson1982a, Zhao2014, Marzari1997, Taherinejad2014,
Li2015k, Gresch2016, Liu2016q, Alexandradinata2016, Lian2017} . If there is only one band
or counting all bands as a whole (which is the case when the Fermi energy is below or
above all the bands), the project operator retrogresses to identity matrix or to zero
matrix and the topological number must take trivial value, i.e., the total Zak phase of
two bands in 1-dimensional (1D) Su-Schrieffer-Heeger (SSH) model is $2n \pi$ ($n$ is a
integer)\cite{Su1979, Su1980, Kivelson1982, Li2014a, Rhim2017} and the Chern number of the
entire bands on a 2-dimensional (2D) lattice must be $0$ \cite{Thouless1982}. In this
article, the above statement can be violated when the non-hermitian Hamiltonian is
intrinsically defective in the BZ. The eigenstates will only cover a subspace of the
Hilbert space\cite{RevModPhys.93.015005}, which is referred as the vacuum of band in this
article.

\begin{figure}[ht] 
  \includegraphics[width=0.38\textwidth]{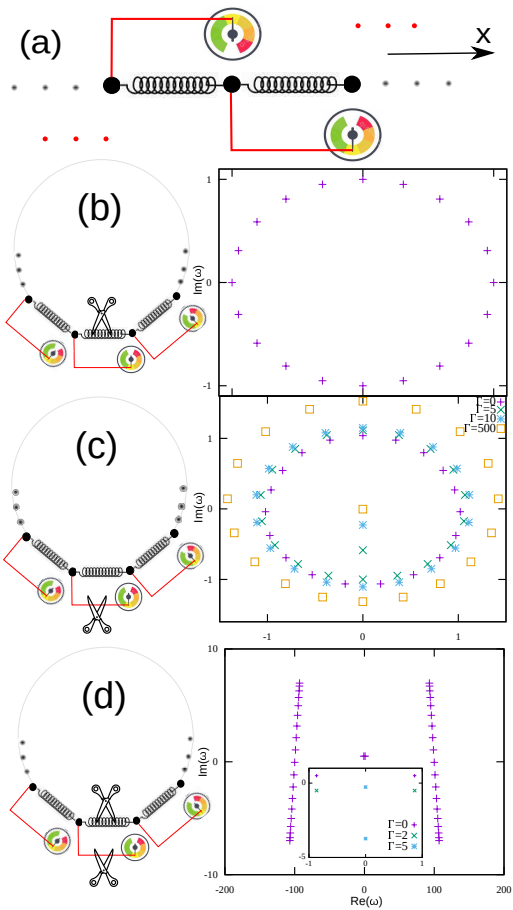}
  \caption[Fig]{\label{fig1} (a) MPs are connected with their nearest neighborhoods by
	  springs whose spring constant $K=1$. The ACs, $x_i-\dot{x}_{i+1}=0$, rigidly
	  bond the displacement of the $i$th MP with the velocity of the next MP. Each MP
	  feels the restoring force, the stress force $\lambda_i$ from the bond and the
	  frictional force $-\Gamma \dot{x}_i$. The length of the chain is $L=20$. To
	  check the calculation, we may also replace the stress force $\lambda_i$ by the
	  restoring force from a strong spring, $\lambda_i \sim -K' (x_i -
	  \dot{x}_{i+1})$, where $K'=10^4$ in the calculations.  (b) The spectrum when one
	  spring is removed. There is no boundary state. (c) The spectrum when one
	  constrain is removed. One boundary state emerges in this case. Its frequency is
	  mixed with those of bulk states when $\Gamma=0$. But when increasing $\Gamma$,
	  it approaches to the original point.  (d) When both spring and constrain are cut
	  off, the spectrum is largely modified by the skin effect. The insert shows the
	  spectrum near the center of complex plane. The topologically protected boundary
  state still survives. }
  \end{figure}

Our models are classical vibration lattices made up by mass points (MPs) connected by the
springs.  The key ingredient is the active constrains (ACs) subjected on each MPs.
Compared with the passive constrains, the ACs need extra steps such as measurement and
analysis. For instance, in Fig.  \ref{fig1}(a), we show the ACs with many measurement
instruments. Each MP is rigidly bonded with the pointer of speedometer on the next MP. In
the second 2D model, we need more complicated AC instruments because each MP will be
bonded with the pointer of a machine analysis whose output is determined by the movements
of several neighboring MPs. 

  In the first model, we find a chain with only one band. But its Zak phase is $\pi$.  As there is
only one band contributes, only one boundary state is topologically protected.  It will
stay at either right or left end of the chain. In the absence of the chiral symmetry, the
vibration frequency of this boundary state is not $0$. But when the frictional
forces dominated, an effective chiral symmetry restores. So an interesting phenomenon
emerges: as subjecting stronger frictional forces, the damping rate of the boundary state
is less.  In the second model, we study a 2D lattice with two bands. Each band contributes
the Chern number $1$ so that the Chern number of the whole bands is $2$ instead of $0$.
The boundary states will not always connect different bands continuously. But in order to
compensate the pumped bulk states due to the nonzero Chern number, the boundary state must
have a moment to spread to the bulk. So another abnormal behavior appears: the in-gap
states are topologically protected to extend to the bulk.

{\it 1D lattice model.---} The 1D classical lattice is made up by identical MPs with msss
$m=1$. Only the movements in the direction along the chain are considered.  When ACs are
applied, one band will be frozen and only one band persists. After a Fourier
transformation, the constrains become $x_k -e^{Ik} \dot{x}_k =0$.  So the states on the
vibration band must be 
\begin{equation}
	| \psi_k \rangle \equiv \begin{pmatrix} x_k \\ \dot{x}_k \end{pmatrix} = 
	\frac{1}{\sqrt{2}}\begin{pmatrix} 1 \\ e^{-Ik} \end{pmatrix}.
\end{equation}

One should be careful in applying the Lagrange multiplier $\lambda$. The variational term
$\dot{\lambda}$ does not appear in the equation of motion. This is because when representing $\lambda_i$ as the
stress force induced by the $i$th bond, the reaction force in the bond does not act back
to the $(i+1)$th MP but to the pointer of the speedometer.  After appending the Fourier
transformation of $\lambda_i$ to the MP's kinetic state $|\psi_k \rangle$, the equation of
motion for $|\psi'_k \rangle \equiv (x_k, \dot{x}_k,\lambda_k)^T$ is
\begin{equation}
	\begin{pmatrix} 1 & 0 & 0 \\ 0 & m &0 \\ 0 & 0 & 0 \end{pmatrix} \dot{|\psi'_k
	\rangle} =
	\begin{pmatrix} 0&1&0 \\ -4K\sin^2(k/2)& -\Gamma & 1 \\ 1& -e^{Ik}&0 \end{pmatrix}
	|\psi'_k \rangle.
\end{equation}

After substituting $|\psi'_k \rangle = e^{-I\omega t} |\psi'_k(\omega) \rangle$, the
eigen-frequency $\omega$ and the eigenstate $|\psi'_k(\omega) \rangle$ are found by
solving a general eigen-problem likes, $ -I\omega C |\psi'(\omega)\rangle = D
|\psi'(\omega) \rangle$, where $C$ is an irreversible matrix.  As $D$ is reversible, by
multiplying $-\frac{D^{-1}}{I\omega}$ from the left, the above equation is equivalent to
the eigen-problem of a non-hermitian matrix $D^{-1}C$, $-\frac{1}{I\omega}
|\psi'_k(\omega) \rangle = D^{-1}C | \psi'_k (\omega) \rangle $. The matrix must be right
at exceptional point (EP) \cite{Moi, Berry2004, Heiss2012, Mehri-Dehnavi2008, Liang2013,
Malzard2015, Cerjan2016, Lin2016a, Xu2016e, RevModPhys.93.015005} because, physically,
constrains can only freeze vibration modes instead of introducing new ones. For the
present 1D model, there are only two eigenvalues, $-\frac{1}{I\omega} = e^{Ik}$ and $0$,
where the first one is associated with the vibration band $|{\psi'}^{(1)}_k(\omega)
\rangle$ and the latter is two fold degenerated with only one corresponding eigenstate (in
the frozen band), $|{\psi'}^{(2)}_k (\omega) \rangle$. The part of the eigenstate that
describes the motion of the MPs, $|\psi^{(i)}(\omega)\rangle$, is obtained by eliminating
the stress variable $\lambda_k$ in $|\psi'\rangle$. The operator $\sum_k |
\psi^{(1)}_k(\omega) \rangle \langle \psi^{(1)}_k(\omega)|$ cannot span the whole phase
space and is not identical matrix.  Evidently, such operator does not commute with
$e^{I\hat r \delta k}$ so that the single vibration band can have nontrivial topological
phase. This problem is more clear when the AC bonds are replaced by strong springs.  One
immediately realize that the ACs are freezing a band and pushing their frequencies to
$\infty$.  The Zak phase of the frozen band is also $\pi$. So the problem can be read as:
the visible band is left on the remnant of the frozen band.

Actually, the operator $\sum_k |\psi^{(1)}_k(\omega) \rangle \langle
\psi^{(1)}_k(\omega)|$ is no longer a project operator on the eigenstates. This leads
to the debate on how to define the topological phase in non-hermitian systems. If one
insists to define the topological phase from the project operator, the Zak phase should
be $I\int_0^{2\pi} dk  \langle \langle \psi'^{(1)}_k | \frac{\partial}{\partial k}
|\psi'^{(1)}_k \rangle$, where $\langle \langle \psi'^{(1)}_k|$ are the left eigenstates of
the matrix $D^{-1}C$. But this expression includes the contributions from the stress forces.
The forces are extra variables
which are introduced to complete the equation of motion. They have nothing to do with the
phase space of the MPs. Furthermore, in principle, we can introduce similar auxiliary variables
as many as possible, e.g., employing more variables like $\alpha_k=\beta_k=\lambda_k$ and
appending all of them to $|\psi_k \rangle$. So there is no reason to count their contributions to
the topological phase. Besides that, the above Zak phase is not real in general. If one
would like to represent the topological phase back to the relative position of the
Wannier function's center in each unit cell, a complex position value is not acceptable.
So in this article, we take the Zak phase as
\begin{equation}
	I\int_0^{2\pi} dk \langle \psi^{(1)}_k| \frac{\partial}{\partial k} |\psi^{(1)}_k \rangle.
\end{equation}

In Fig. \ref{fig1} (b), (c) and (d), we show the vibration spectrum for a ring with
different boundary conditions. A open boundary-like condition can be introduced by: (b)
cutting off one spring , (c) cutting off one constrain and (d) cutting off both spring and
constrain.  In the calculations, we have used the trick by replacing all rigid constrains with
the strong springs. The rigid constrains are restored when $K' \to \infty$. 

In Fig. \ref{fig1}(b), with only one spring is cut off, the spectrum is same as that of
the perfect ring. This is because the constrains keep restricting the possible vibration
modes in the subspace that is identical to the perfect ring. The lost of spring cannot
modify the spectrum. 

In Fig. \ref{fig1}(c), when the constrain between the ends is removed, the bulk spectrum
is similar to that of the perfect ring. But there is one boundary state at the right end
of the chain. This single topologically protected boundary state can be understood by
recalling the arguments in the electronic SSH model because their bulk band states share
the same equation. In the topological SSH model, there are two boundary states. Each one
is contributed from the up and the low bands half by half.  But in the present model, only
one band contributes so that there is only half boundary state at each end. But as half
states are unstable, they recombine to one state.  We also find that its frequency is
not zero when $\Gamma$ is small and is on the imaginary axis. This is because there is no
chiral symmetry in the non-hermitian Hamiltonian. But the equation of motion and the ACs
are invariant under complex conjugation. So the frequency of the boundary state must stay
at the imaginary axis because of $\omega \to -\omega^*$ symmetry.  When the frictional
forces dominate, the equation of motion approaches to $\ddot{x_i}=-\Gamma \dot{x_i}$.
Under time reverse operation, this equation changes sign and the ACs become $x_i
+\dot{x}_{i+1}=0$. So after combining the time reversed model and the model itself, there
are totally two bands and these two bands will mimic those in the SSH model. This means
that an effective chiral symmetry restores in the combination and the frequency of the
boundary state must be fixed at zero. This is why the damping rate of the boundary state
decreases when $\Gamma$ increases.

\begin{figure}[ht] 
  \includegraphics[width=0.45\textwidth]{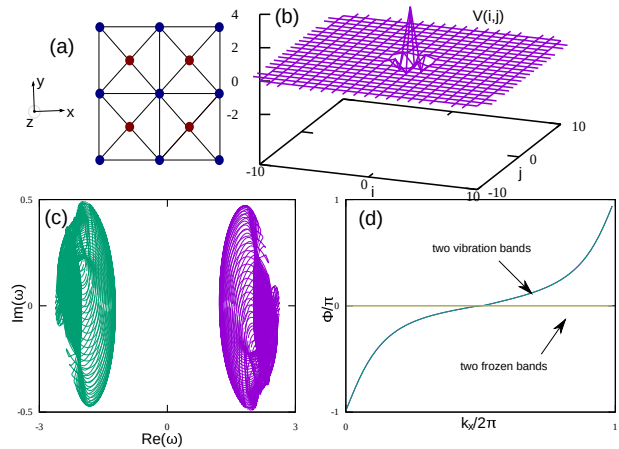}
  \caption[Fig]{\label{fig2} (a) The 2D square lattice. The blue and red circles represent
	  type A and type B MPs and the black lines are the springs. All springs are taut
	  so that they can generate restoring forces when the MPs are moving in the $z$
	  direction (perpendicular to the plane). We take the spring constants 
	  $K_1=K_2=0.5$. The ACs on the B type MPs are not shown explicitly. (b)
	  The value of $V(i,j)$. (c) The spectrum of the two vibration bands on
	  the complex plane. (d) $\Phi$ as the function of $k_x$. It indicates that the
  Chern number of the two vibration bands is $1$, respectively. While the two frozen bands have the
  Chern number $0$.}
\end{figure}

In Fig. \ref{fig2}(d), the spectrum is entirely different from that of the perfect ring.
All bulk states are squeezed to the ends and their frequencies are largely modified as
well. This is the skin effect in non-hermitian
system\cite{Xiong2018a,Weidemann2020,Gao2020}.  We also found that this skin effect does
not obey the Theorem I in Ref. \onlinecite{Okuma2020} because the Hamiltonian is
defective.  Near the center of the complex plane, there are two states.  One is originated
from the topologically protected boundary state and the other is from the vibration mode
released by the missing constrain. They hybrid together and can be separated by the
frictional forces in bulk.

We know there are two equivalent ways to study topological physics in the hermitian
systems: one is on the Hamiltonian, especially on the symmetry of the Hamiltonian and the
other is on the Berry phase. One of our motivations is to find what will happen when this
equivalence breaks down: there are topological-like eigenstates but the non-hermitian
Hamiltonian does not have the corresponding symmetry. The above model shows that the
bulk-boundary correspondence still holds. But the number of the boundary states and their
frequencies are different from those of the hermitian case. Our next model is to find what
will happen when this idea is applied to the 2D Hall system.

{\it 2D Hall model.---} The 2D model is shown in Fig. \ref{fig2}(a). Here we only study
the transverse-mode of the wave.  Each unit cell has two MPs (denoted as A and B types of
MPs) with the identical mass $m=1$.  The ACs are only acted on the B type of MPs by the
restriction
\begin{equation}
	B_{\vec{k}} -G(\vec{k},A_{\vec{k}},B_{\vec{k}})=0,
	\label{eq2_1}
\end{equation}
where
\begin{multline}
	G=-[\sin(k_x)+I\sin(k_y)] A_{\vec{k}} +[\cos(k_x)
+\cos(k_y) \\ 
+\sqrt{(1-\cos(k_x)-\cos(k_y))^2+\sin^2(k_x)+\sin^2(k_y)}] B_{\vec{k}}.
\end{multline}
Here $\vec{k}$ is the wave vector
and $A_{\vec{k}}$($B_{\vec{k}}$) is the Fourier's transformation of
the displacements for the A(B) type MPs. Such constrains can be manipulated by the rigid
connections between the B type MPs and the pointers of the machines whose outputs are
$G_{i,j}$, where $G_{i,j}$ is the inverse Fourier's transformation of
$G(\vec{k},A_{\vec{k}},B_{\vec{k}})$,
\begin{multline}
	G_{i,j}=\frac{1}{2}[I(A_{i+1,j}-A_{i-1,j})+(A_{i,j-1}-A_{i,j+1})+ (B_{i+1,j} \\
	+B_{i-1,j}+B_{i,j+1}+B_{i,j-1}) + \sum_{i_1,j_1} V(i_1,j_1) B_{i-i_1,j-j_1} ].
	\label {eq2_2}
\end{multline}
Here $A_{i,j}$($B_{i,j}$) is the displacement of the A (B) type MPs at the site $(i,j)$,
$V(i_1,j_1)$ is the inverse Fourier's transformation of
$2\sqrt{(1-\cos(k_x)-\cos(k_y))^2+\sin^2(k_x)+\sin^2(k_y)}$ and the summation is over all
2D lattices. As shown in Fig. \ref{fig2} (b), $V(i,j)$ decreases rapidly when $i$ and $j$
are away from the origin. The imaginary number $I$ in Eq.  \ref{eq2_2} can be realized by
a $\pi/2$ phase shift in the machines.

We have calculated the vibration spectrum using the same trick introduced in the 1D case.
The ACs freeze two bands and leave us with another two bands. They are plotted on the
complex plane in Fig. \ref{fig2} (c).  We also calculate the Chern number of each band by
watching the Wannier function's center as the function of $k_x$, $\Phi(k_x)=\int dk_y
\langle \psi_n(k_x,k_y) | I \frac{\partial}{\partial k_y} |\psi_n(k_x,k_y)\rangle$, where
$|\psi_n \rangle$ is the kinetic state of the MPs for the $n$th band.  As Fig.
\ref{fig2}(d) shows, the Chern number of each band is $1$. The figure also implies that
the Chern numbers for the two frozen bands are $0$. Interestingly, even after counting the
contributions from the frozen bands, the total Chern number is $2$ instead of $0$. This is
different from that in the 1D case.

\begin{figure}[ht] 
  \includegraphics[width=0.45\textwidth]{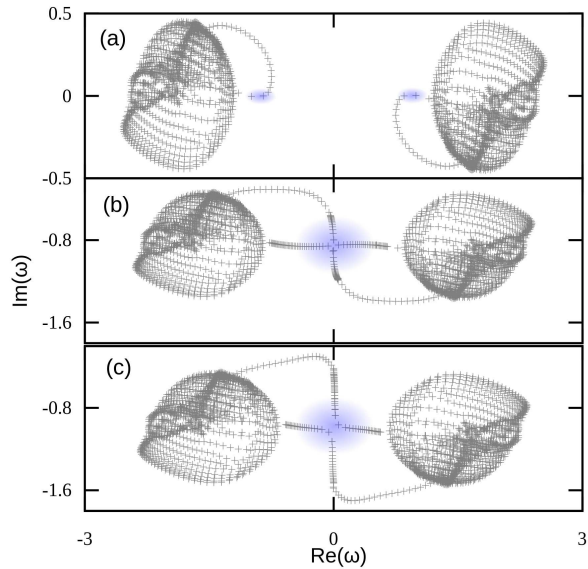}
  \caption[Fig]{\label{fig3} The spectrum for a 2D stripe. The width is $W=20$. The data
	  of the boundary states within the gap are closely spaced for the sake of clarity
	  (the data are pasted from the calculation with higher density $k_x$). The spring
	  constant and other parameters are the same as those in Fig. \ref{fig2}. The
	  frictional coefficient is (a) $\Gamma=0$, (b) $1.7$ and (c) $2$. The blue
  regions indicate that the states in them are extended in bulk.}
\end{figure}

Finally, the edge states are studied. Like that in the 2D quantum Hall effect, we consider
an infinitely long stripe. It is placed along the $x$ axis so that $k_x$ is good number.
All connections, including the springs and the components of the ACs, those across the
edges are cut off. As Fig. \ref{fig3} shows, unlike that in the quantum Hall effect, the
boundary states do not link the two bands when $\Gamma$ is small. Also the open boundary
condition does not modify the bulk spectrum.  So the nonzero Chern number of the bulk
bands should topologically protect the existence of the boundary states. These two
phenomena seem controversial when adopting the discussion from the quantum Hall effect: as
$k_x$ is varying by $2\pi$, one bulk state is pumped from one edge to the other and the
energies of the edge states must cross over the gap to compensate this extra pumped state.
Otherwise, the system is not periodic along $k_x$ anymore. But our data cleanly illustrate
a different picture. It is more like a situation (not really exist) in quantum Hall effect
that the neighboring bands disappear and spectrum of the edge states are connected by
themselves. But in quantum Hall effect, they cannot be connected because the states are on
the different edges. The above argument implies us to check the eigen-states of these
boundary states.  Interestingly, we find that they do become extended in the blue regions.
So the situation changes to: as $k_x$ is varying by $2\pi$, one bulk state is pumped to
the edge while the boundary states dissipate the state back to the bulk. In doing so, the
boundary states in the gap need to be extended in bulk at sometime. This is the
topological reason behind the in-gap extended states.

{\it Discussions.---} Due to the skin effect in non-hermitian systems, it is known that
the bulk-boundary correspondence cannot be applied directly because the topology of bulk
states in the $k$ space is different from that of skin states in the presence of open
boundary\cite{Silveirinha2019,PhysRevB.99.201103,PhysRevB.102.085151,Helbig2020,zhang2021tidal}.
But in most cases of this article, the bulk spectrum does not change dramatically when the
boundaries are introduced. So our discussions on the correspondence between boundary
states and topology of the bulk states are still valid. 

The models we presented here are different from the classical isostatic lattices proposed
by Kane and coworkers\cite{Kane2013,Meeussen2016}, where the passive constrains are
subjected to the lattice and freeze all bulk vibration modes.  Mostly, the topological
index is winding number in any dimension and the topological boundary states are zero
frequency modes. While in the above models, it is the ACs that split the Hilbert space
into two twisted subspaces, the allowed and the forbidden subspaces. The bulk vibration
states can still survive in the allowed subspace. The topological index is Zak phase,
Chern number and others.  The models are also different from the topological acoustic
structures induced by strain\cite{Abbaszadeh2017,Yang2017,Frenzel2017,Chaunsali2017}, or
Coriolis force\cite{Wang2015j,Wang2015w,Xiong2016}.

The models imply that the topology of a band can be richer than those in the hermitian and
the undefective non-hermitian systems. When the whole BZ becomes exceptional, the
Hamiltonian and its symmetry cannot uniquely determine the topology and one needs to look
back upon the truncated wave-functions.  The aggregate of topological systems is greatly
enlarged in this case.

{Acknowledgments.---} 
The work was supported by 
National Foundation of Natural Science in China Grant Nos. 10704040.

\bibliographystyle{apsrev4-1}
\bibliography{main}

\end{document}